\documentclass[aps,prb,twocolumn,floatfix,groupedaddress]{revtex4}

\usepackage{graphicx}
\usepackage[usenames,dvipsnames]{xcolor}
\usepackage{enumitem}

\usepackage{latexsym}
\usepackage{amsmath}
\usepackage{dcolumn}
\usepackage{bm}
\usepackage{graphicx}
\usepackage{epsfig}
\usepackage{verbatim,listings}
\usepackage{multirow}
\usepackage{dcolumn}

\newcommand{\tbh}[3]{\multicolumn{#1}{#2}{#3}}
\newcommand{\tbhc}[1]{\multicolumn{1}{c}{#1}}
\newcommand{\tbhl}[1]{\multicolumn{1}{l}{#1}}

\newcolumntype{f}[1]{D{.}{.}{#1}}

\graphicspath{{figures/}}

\begin{document}                  



\title{Advanced calculations of x-ray spectroscopies with FEFF10
and Corvus}




\author{J. J. Kas}
\affiliation{Department of Physics, University of Washington, Seattle, WA 98195, USA}
\affiliation{Theory Institute for Materials and Energy Spectroscopies,
SLAC National Accelerator Laboratory, Menlo Park, CA 94025, USA}
\author{F. D. Vila}
\affiliation{Department of Physics, University of Washington, Seattle, WA 98195, USA}
\affiliation{Theory Institute for Materials and Energy Spectroscopies,
SLAC National Accelerator Laboratory, Menlo Park, CA 94025, USA}
\author{C. D. Pemmaraju}
\affiliation{Theory Institute for Materials and Energy Spectroscopies,
SLAC National Accelerator Laboratory, Menlo Park, CA 94025, USA}
\author{T.~S. Tan}
\affiliation{Department of Physics, University of Washington, Seattle, WA 98195, USA}
\author{J. J. Rehr}
\affiliation{Department of Physics, University of Washington, Seattle, WA 98195, USA}
\affiliation{Theory Institute for Materials and Energy Spectroscopies,
SLAC National Accelerator Laboratory, Menlo Park, CA 94025, USA}











\begin{abstract}
The real-space Green's function code FEFF has been extensively
developed and used for calculations of x-ray and related spectra,
including x-ray
absorption (XAS), x-ray emission (XES), inelastic x-ray scattering,
and electron energy loss spectra (EELS). The code is particularly
useful for the analysis and interpretation of the XAS fine-structure
(EXAFS) and the near-edge structure (XANES) in materials throughout
the periodic table.  Nevertheless, many applications, such as
non-equilibrium systems, and the analysis of ultra-fast pump-probe 
experiments, require extensions of the code including finite-temperature
and auxiliary calculations of structure and vibrational properties. 
To enable these extensions, we have developed in tandem, a new version FEFF10,
and new FEFF based workflows for the Corvus workflow manager, which allow users to
easily augment the capabilities of FEFF10 via auxiliary codes.
This coupling facilitates simplified input and automated calculations of spectra
based on advanced theoretical techniques. The approach is illustrated with
examples of  high temperature behavior, vibrational properties,
many-body excitations in XAS, super-heavy materials, and
fits of calculated spectra to experiment.
\end{abstract}

\maketitle                        


\section{Introduction}

The real-space Green's function (RSGF) code
FEFF \cite{rehralbers00,feffcr,FEFFPCCP} has been extensively developed
and is in wide use for calculations of x-ray absorption (XAS) and a number of
related spectroscopies \cite{vanbokhoven}.
Owing to its versatility and broad applicability, FEFF has become a workhorse
for simulations and analysis of extended x-ray absorption fine-structure
(EXAFS). However, both the theory and experiment continue to increase in
sophistication.  For example, the advent of time-domain spectra 
at XFEL sources and increased resolution have enabled many
novel experimental studies. Improved treatments of many-body effects,
such as inelastic losses and thermal vibrations, have also been developed.
As a consequence, extensions to the computational framework are
desirable. At the same time, it is no longer efficient 
to include all of these extensions within a single monolithic code like
FEFF9 and its predecessors.
For many advanced or high-quality calculations, input from auxiliary codes
based on DFT, quantum-chemistry, and Monte Carlo sampling methods are needed. 

To address these issues, we have developed a hybrid computational approach
based on a new version FEFF10 and the recently developed Corvus workflow
tool \cite{Story2019}. 
The monolithic FEFF10 code implements several
extensions, while Corvus enables workflows that link
multiple codes seamlessly to facilitate advanced and custom
applications. In Sec.\ \ref{sect:rsgf} we describe the real-space
Green's function approach implemented in FEFF10, and
in Sec.\ \ref{sect:corvus}, a brief summary of the Corvus workflow
framework together with examples of advanced calculations
that can be implemented with 
 FEFF10 and Corvus. In particular, we discuss
finite-temperature simulations, vibrational effects, inelastic losses
and multi-electron excitations, spectra of superheavy elements,
full spectrum optical constant, and fitting of
theoretical XANES spectra to
experimental data. Finally, Sec.~\ref{sect:conclusion} 
summarizes the status of the Corvus/FEFF10 capabilities and possible
improvements.


\section {RSGF Theory of X-ray spectra \label{sect:rsgf}}


 The RSGF theory of x-ray spectra is now well established,\cite{rehralbers00}
and has been automated and optimized in the several generations of FEFF.  
In contrast to traditional wave-function approaches, the RSGF approach
is based on a real-space treatment of the one-particle Green's 
function. This strategy is advantageous, since it avoids 
calculations of and sums over eigenstates
as in the traditional Fermi's golden rule approach.
The approach also facilitates the inclusion of key many-body effects such as inelastic losses.
The theory and can also be generalized straightforwardly to
finite temperature (FT) \cite{Tun21}.
 The implementation of the theory in FEFF is modular and designed to facilitate systematic improvements.  Consequently,
extensions described here utilize much of the code base in
  previous versions \cite{rehralbers00,feffcr,FEFFPCCP},
and permit backward compatibility.  Thus, we only summarize the basic
elements of the theory here, and focus 
on the extensions included in FEFF10 and the Corvus interface
with auxiliary codes in the next sections.

A key approximation in the RSGF formalism
is the {\it muffin-tin approximation},
in which the scattering potential $v(r)$ is partitioned into
Voronoi cells and approximated as spherical within each cell,
$v(r)=\sum_R v_R(r)$. Then the Green's function can be solved
exactly in terms of radial wave functions and spherical harmonics.
The contribution to the XAS from a given 
core level is then expressed compactly as
\begin{equation}
\label{Lxas}
\mu_1 (E) 
= {4\pi^2} \frac{\omega}{c}\,
\sum_{L,L'}M_{L}^*\ \rho_{L,L'}(E+E_c-\mu)M_{L'}[1-f(E+E_c-\mu)],
\end{equation}
where $M_{L}$ is the dipole matrix element between the core-state and a
scattering-state $| L \rangle$ 
of angular momentum $L$ at the absorbing atom site $R=0$,
and 
$\rho_{L,L'}(E) = -(1/\pi) {\rm Im}\, G_{L,L'}(E)$
are matrix elements of the density-matrix spectral function at
the absorbing atom $R=0$, where for simplicity the site indices
$R$ and $R'$ have been suppressed. Finally, $E_c$ and $\mu$ are
the energy of the core-level and the chemical potential, and $f(E)$ is
the Fermi function.
The dipole matrix elements $M_{L}$ couple the scattering
(photoelectron) states to
relativistic atomic states, calculated using an
automated single configuration version of 
the multi-configurational
Dirac-Fock atomic code of Desclaux \cite{DESCLAUX1973311,Ankudinov_1996},
which we have recently extended to treat superheavy elements up to
$Z=138$ \cite{Zhou2017}.
Next the Green's function $G$ is separated into central atom and
scattering parts $G=G_c + G_{sc}$, which are calculated separately.
The central atom part can be
represented in a basis of relativistic, spherically symmetric
scattering states $|L,R\rangle =R_L(r)Y_L(\hat r)$.
The scattering part can be expressed in terms of the free Green's
function $G_0$ and scattering $T$-matrices at each site,
\begin{equation}\label{fmsg}
G^{\rm sc}_{L,L'}=\left[(1-\bar G^0 T)^{-1}\bar G^0\right]_{L,L'}.
\end{equation}
where the $T$-matrix is diagonal in $l$ and $R$,
$T_{L,L',R,R'}=t_l \delta_{l,l'} \delta_{R,R'}$,
$t_l = \exp(i\delta_l)\sin(\delta_l)$,
$\delta_l$ are the scattering phase shifts \cite{rehralbers00}, and $\bar G^0$ is $G^0$ with the diagonal elements set to 0 to avoid scattering from the same atom consecutively.
For the XANES region (where $l_{max}$ is typically about 4), the matrix
inverse is easily calculated, i.e., by full-multiple-scattering (FMS).
For the EXAFS where $l_{max} \sim 20$, inversion is
computationally prohibitive, and it is more efficient
to expand the matrix inverse in a rapidly converging
geometric series, corresponding to the MS {\it path expansion},
\begin{equation}\label{msg}
  G^{sc}_{L,L'} =  
{\left[\bar G^0 + \bar G^0 T \bar G^0 + \bar G^0 T \bar G^0 T \bar G^0 + \cdots \right]}_{L,L'}.
\end{equation}
Here the successive terms represent single, double, and higher order
scattering processes, which are calculated using a separable
approximation \cite{rehralbers00}.
Conventionally the XAS is expressed in terms of
a quasi-atomic background $\mu_0$ from the central absorbing atom, and a
scattering part $\mu_1 =\mu_0(1+\chi)$.
Using the path expansion, the {\it fine structure} $\chi$ due to
MS of the photoelectron by
the environment can be expressed in terms of the EXAFS equation, like that of 
Sayers, Stern and Lytle \cite{sayers71},
\begin{equation}
\label{eq:EXAFS}
\chi(k)= S_0^2\,\sum_{R} \,\frac{|{f_{\rm eff}(k)}|}{kR^2}
\sin(2kR+\Phi_k) e^{-2 R/\lambda_k} e^{-2\sigma ^2 k^2}.
\end{equation}
It is important to note that the scattering amplitudes
$f_{\rm eff}(k)$ (from which the FEFF code takes its name)
include important curved wave corrections. The same representation
 applies to
both single- and multiple-scattering contributions where $f_{\rm eff}$ is
defined for a given path. The EXAFS equation
also includes path and temperature dependent Debye-Waller factors
$e^{-2\sigma ^2 k^2}$ due to fluctuations in the interatomic
distances $R$. 

Corrections for multi-electronic excitations can be included in terms of 
a convolution over the core-spectral function $A_c(\omega)$.
\begin{equation}
\label{eq:mumb}
  \mu(\omega)=\int_0^{\infty} \!d\omega'
 \mu_1(\omega') {A_c} (\omega-\omega')\equiv \langle \mu_{qp}(\omega)\rangle,
\end{equation}
where $\omega'$ is the excitation energy~\cite{campbell02}.
Formally the core spectral function
$A_c(\omega)= \Sigma_n |S_n|^2 \delta(\omega-\epsilon_n)$, where $S_n = \langle \Phi_0^{N-1}|\tilde \Phi_n^{N-1}\rangle$ is a many-body overlap integral.
This function characterizes the effects of inelastic losses and
leads to an energy-dependent broadening of the XAS.
 Likewise, the net EXAFS is also given by a convolution
with the quasi-particle fine structure. For each
MS path $R$, this convolution leads to an amplitude reduction factor
$S_R^2(\omega)$ which is roughly constant, and a negative phase
shift $\Phi_R(\omega)$,
$\langle e^{2ikR} \rangle  = S_R^2(\omega)\, e^{2ikR + \Phi_R(\omega)}.$
Typically $S_0^2 \approx 0.9$, consistent with
the reduction observed in EXAFS experiment \cite{campbell02}.
Approximate calculations of the core-spectral function $A_c(\omega)$
in Eq.\ (\ref{eq:mumb}) are  also possible in FEFF, using an atomic
approximation with the SFCONV card.
 However more precise calculations require auxiliary calculations,
as discussed in Sec.\ \ref{sect:mbconv}.

\section{Corvus and FEFF10 \label{sect:corvus}}

FEFF10 has a variety of extended capabilities for calculations of electronic
structure and spectra. However, for many advanced or high
quality calculations, FEFF must be augmented with calculations of 
structure and other properties based on DFT, quantum-chemistry,
Monte Carlo sampling, etc. Examples include
calculations of spectra based on DFT optimized structure, inclusion of
vibrational effects through calculations of the dynamical matrix \cite{dmdw,frontiers1,frontiers2} or
molecular dynamics simulations \cite{PRB.2012.85.24303,TJoPCL.2017.8.3284,PRB.2008.78.121404}, or treatment of multi-electron excitations
through real-time TDDFT. For these purposes, an intelligent workflow tool dubbed Corvus
has been developed \cite{Story2019}. This tool replaces JFEFF, the Java-based
GUI of FEFF9, which is difficult to maintain and extend.   Instead,
the combination of FEFF10 and Corvus facilitates 
many advanced calculations previously limited to expert users.
In this section we briefly describe Corvus, and give details on the
extensions to FEFF available within the 
FEFF10  code itself, as well as in combination with Corvus,
which is now 
the preferred method for running FEFF10.

Briefly, Corvus is a Python-based workflow framework which consists of several layers of code. First a set of general workflow tools and an internal system dictionary that holds a description of all of the scientific properties of the physical system at hand provides the capability to easily develop, or automatically generate workflows. Second, external software facing ``handlers'' translate between the format of the internal system dictionary and specific external scientific software packages. In addition, these handlers provide a layer of automation in terms of smart, input derived default parameters, as well as error catching capabilities. Because the handlers translate all input and output to a standard Corvus format, and automate many of the code specific parameters associated with specific external software packages, the input is relatively simple, allowing the user to focus mostly on the physical system at hand.

In addition to the Python workflow framework, we have developed
several Corvus web interfaces. The first serves as a demonstration of
the capabilities of FEFF10 and Corvus, and allows anyone to try a few
simple calculations. It is hosted on the local TIMES cluster at
SLAC \cite{feffTIMESWeb}. The second interface provides
a portal to the  
National Energy Research Scientific Computing Center (NERSC). This
allows users to easily set up, run, and monitor calculations using
Corvus and a variety of underlying scientific software \cite{portal}.

\subsection{Electronic and Lattice Temperature Effects }

With the combination of FEFF10 and Corvus, it is possible to treat
spectra at finite temperature, including electronic and lattice
temperature effects. While the electronic
temperature effects are built into FEFF10, the treatment of
FT lattice effects generally requires auxiliary calculations.
The combination of FEFF10 and Corvus allows for the simulation of
a broad range of temperatures from zero to the warm dense matter regime. In
addition, this allows for simulations of non-equilibrium states of matter,
such as those produced during pump-probe and shock experiments. Here we
briefly summarize the treatment of finite temperature effects, which are 
discussed in more detail elsewhere \cite{Tun21}. 

One of the new options in the FEFF10 code is the inclusion of a
finite temperature (FT) generalization of the RSGF algorithms.
The effects of temperature on the electronic system require
extensions of several parts of the theory \cite{Tun18}.  First, the
self-consistent field (SCF) calculations have been
updated to include Fermi-Dirac occupations by integration in the
complex plane, and summing over the appropriate Matsubara
poles. Second, the exchange correlation potential in the SCF has
been updated to include explicit temperature
dependence \cite{Trickey14,Trickey16}.
Third, Fermi-Dirac statistics are included.
At low temperatures the effects of temperature on the electronic
system cause the chemical potential to shift following the
Sommerfeld expansion, while the edge broadens and lowers
 due to the Fermi-Dirac statistics. At higher
temperatures, however, the electronic structure changes, and the shift
in the chemical potential and shape of the XAS can deviate
from the Sommerfeld approximations. Finally, a temperature dependent
self-energy can be important at very high $T$.
These observations can be explained by noting that
at low temperatures compared to the Fermi temperature $T << T_F$ (which is typically of order $10^4$ K), the exchange-correlation potential and self-energy are weakly temperature dependent. Thus a zero-temperature approximation is often adequate for electronic structure, although vibrational effects become substantial for $T$ of order the Debye temperature $\theta_D$ (which is typically $10^2$-$10^3$ K). However, in the WDM regime $T \sim T_F$, explicit temperature
dependence is necessary, as the exchange-correlation potential changes from
exchange- to correlation-dominated behavior in the WDM \cite{KasBlantonRehr19}.

The primary effect of the electronic temperature is the inclusion of
Fermi factors for the occupied- and un-occupied levels in the calculations of the cross-section. Notably, these effects lower the edge with increasing temperature as determined by the chemical potential $\mu(T)$ which is  determined self-consistently in the finite-$T$ SCF loop. This is illustrated in Fig.\ 1  for several temperatures. The figure also shows a
complementary reduction in the XAS above the edge.

\begin{figure}
        \includegraphics[width=0.45\columnwidth]{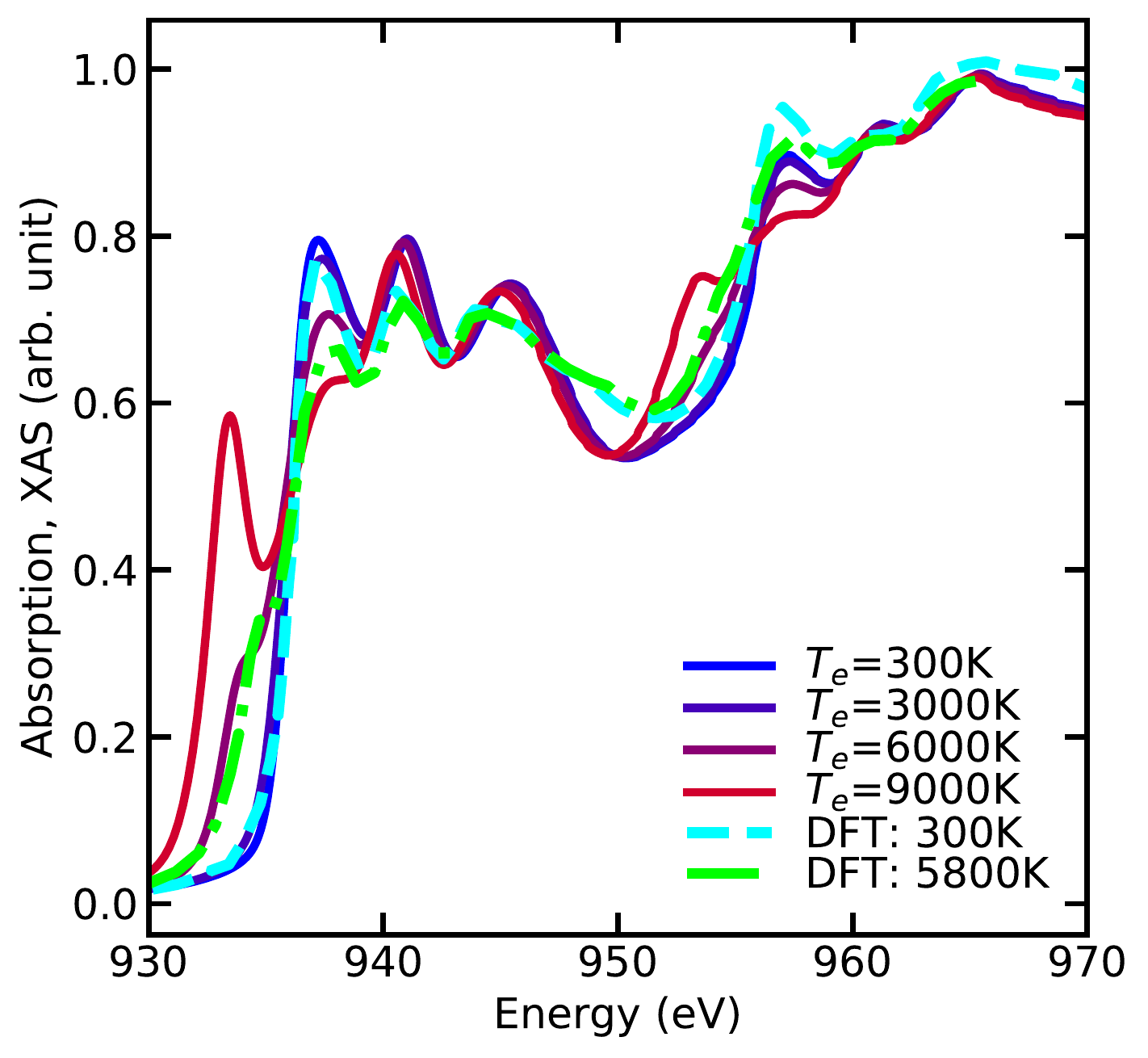}
        \includegraphics[width=0.45\columnwidth]{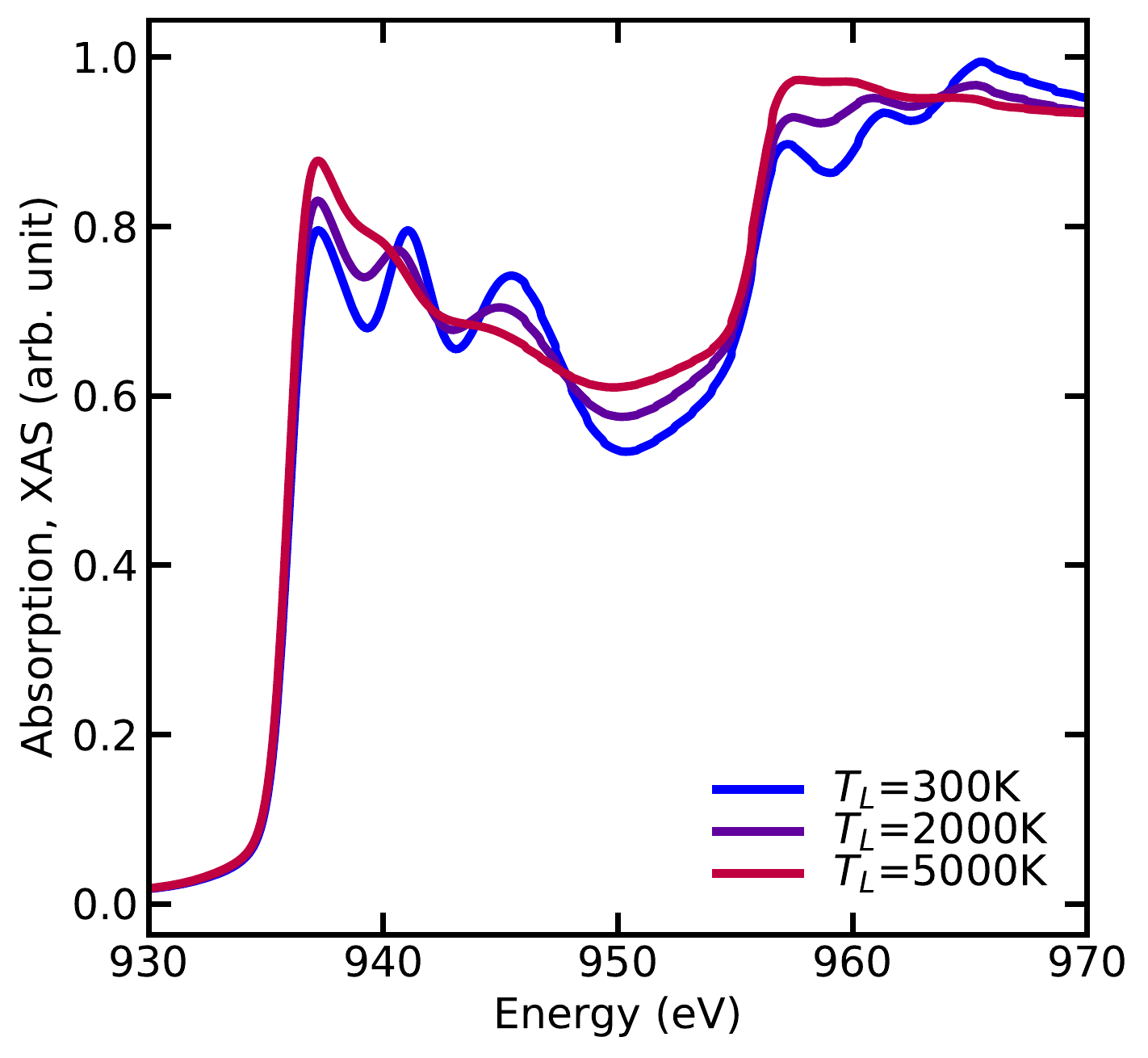}
        \caption{(left) $\textnormal{L}_3$-edge XAS of Cu at normal density for 
electron temperature $T_e=$ 300 K up to 9000 K at lattice temperature $T_L$ = 300 K; (right)   $T_L$ = 300 K up to 5000 K at $T_e$ = 300 K. For comparison, DFT calculations 
        \cite{Jourdain20} are shown for $T_e=$ 300 K (light blue) and 5800 K (green).
}
        \label{fig:cu_xas_temp}
    \end{figure}



\begin{figure}
  \includegraphics[width=0.5\columnwidth,]{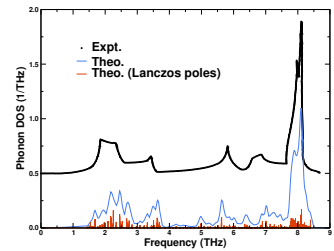}
\caption{Comparion of the experimental \cite{blakemore1982semiconducting} and theoretical total phonon density of states for GaAs. Also shown are the poles generated by the Lanczos algorithm.}
\label{fig:gaaspdos}
\end{figure}

It is essential to treat lattice vibrations in order to describe finite
temperature effects on the XAS accurately. The various regimes
can be treated with different approximations, split by low or
high energy, and low or high temperature.
At low temperatures $T\sim\theta_D$ and
high energies $E-E_0$ above $\sim40$ eV, the spectrum is strongly
damped by lattice vibrations even at zero
temperature through zero-point motion. At low temperatures and
high energies, these effects can be treated via EXAFS
Debye-Waller factors, which damp the fine structure by a
exponential factor $\exp(-2k^2\sigma^2)$ where $\sigma^2$ is the
(path dependent) mean-square relative displacement (MSRD) along the
path in question. There are several methods for approximating the
MSRD, including the use of a correlated Debye or Einstein model, where
the Debye-temperature $T_D$ of the material
can either be treated as a fitting parameter, or estimated
roughly from tabulated elasticity parameters \cite{Anderson1}.
This method is illustrated in Fig.\ 1 (b).
For more complex systems where the correlated Debye model is not
applicable, a more accurate and widely applicable method based on the DFT
calculation of the dynamical matrix can be used \cite{dmdw}.
This method computes the MSRD for a given scattering path $R$ from Debye integrals over the phonon density of states (PDOS) $\rho_R(\omega)$ projected onto that path \cite{poiarkova99-1,poiarkova99-2,krappe02}:
\begin{equation}
\label{eq:dw_fac}
  \sigma_R^2(T) = \frac{\hbar}{2\mu_R}
                  \int_{0}^{\infty}{ \frac{1}{\omega}
\coth \Bigl ( \frac{\beta\hbar\omega}{2} \Bigr ) \rho_R\left(\omega\right)}
  ~ d\omega,
\end{equation}
where $\mu_R$ is the reduced mass for the path and $\beta = 1/k_{B}T$.
In the DMDW module of FEFF10, the PDOS is calculated using a continued fraction representation of the phonon Green's function generated with the iterative Lanczos algorithm \cite{deuflhard95}:
\begin{equation}
\label{eq:greensf}
  \rho_R(\omega) = - \frac{2\omega}{\pi}
\mathrm{Im} \Bigl < 0 \Bigl | \frac{1}{\omega^2-{\bf D}+i\epsilon}
\Bigr | 0 \Bigr >\;,
\end{equation}
where $\left | 0 \right >$ is the Lanczos seed for a
mass-weighted normal displacement of the atoms along the path, and ${\bf D}$ is the dynamical matrix of force constants
\begin{equation}
\label{eq:rs_dm}
  D_{jl\alpha,j'l'\beta} =
  \left( M_{j} M_{j'} \right)^{-1/2} ~
  \frac{\partial^{2} E}{\partial u_{jl\alpha}\partial u_{j'l'\beta}} .
\end{equation}
Here $u_{jl\alpha}$ is the $\alpha = \left\{  x,y,z \right\}$ Cartesian
displacement from the equilibrium position of atom $j$ in unit cell $l$, $M_{j}$
is its mass, and $E$ is the energy of the unit cell. Thus, the only quantity required to obtained \textit{ab initio} MSRDs is the dynamical matrix which, however, can not be computed directly in FEFF. This approach has been applied to study XAFS and crystallographic MSRDs of simple \cite{dmdw} systems, but can also be used for more complex materials with negative thermal expansion \cite{frontiers1,frontiers2}.
To streamline the generation of ${\bf D}$, we have developed Corvus handlers to a variety of codes (e.g. ABINIT \cite{gonze2016}, NWChem \cite{nwchem}) that can be used to automatically generate MSRDs, crystallographic Debye-Waller factors, vibrational free energies and phonon densities of state.
Corvus has the advantage that the complicated process of interfacing the computation of the dynamical matrix to the computation of any of those quantities is done automatically. For example, Fig. \ref{fig:corvusgaas} shows a typical Corvus input for the computation of the phonon DOS of GaAs.
\begin{figure}
\begin{lstlisting}[language=sh,%
                   frame=single,%
                   basicstyle=\linespread{1.0}\scriptsize\ttfamily,%
                   commentstyle=\color{RoyalBlue}%
                  ]
target_list {
total_pdos
}
usehandlers { Abinit Dmdw }
cell_scaling_abc {
10.4339965247 10.4339965247 10.4339965247
}
cell_vectors {
0.0 0.5 0.5
0.5 0.0 0.5
0.5 0.5 0.0
}
cell_struc_xyz_red {
Ga -0.125 -0.125 -0.125
As  0.125  0.125  0.125
}
pw_encut 30.0
abinit.ixc 11
pspfiles {
Ga 31-Ga.LDA.fhi
As 33-As.LDA.fhi
}
nkpoints {
8 8 8
}
nqpoints {
4 4 4
}
dmdw.paths {
2
1 1 3.0
1 2 3.0
}
dmdw.nlanc 96
\end{lstlisting}
\caption{Typical Corvus input file for the calculation of the total phonon density of states of GaAs.}
\label{fig:corvusgaas}
\end{figure}
This input simplifies the complex series of steps required to generate the dynamical matrix in Abinit and convert it to the input that DMDW uses. Fig. \ref{fig:gaaspdos} presents the total phonon density of states computed with this workflow, showing that overall agreement with experiment is qualitatively correct. More importantly, the moments of the distribution and their associated mean frequencies (Table \ref{tab:pdosmoms}) are in excellent agreement with experiment, thus ensuring the accuracy of the Debye integrals described above.

\begin{table}
\caption{
Moments and associated mean frequencies for the total phonon density of state  distribution shown in Fig.\ \ref{fig:gaaspdos}.
}
\label{tab:pdosmoms}
\begin{center}
\begin{tabular} {rf{3}f{5}f{4}f{6}}
\hline
& \tbh{2}{c}{Moment (THz$^\mathrm{n}$)} & \tbh{2}{c}{Mean Freq. (THz)} \\
\tbhl{n} &\tbhc{Theory}&\tbhc{Expt.}&\tbhc{Theory}&\tbhc{Expt.} \\
\hline
 -2        &    0.08     &      0.09  &     3.6   &	3.4   \\
 -1        &    0.23     &      0.25  &     4.3   &	4.0   \\
  1        &    5.65     &      5.55  &     5.7   &	5.6   \\
  2        &   37.87     &     36.91  &     6.2   &	6.1   \\
\hline
\end{tabular}
\end{center}
\end{table}

\begin{figure}           
  \centering
  \includegraphics[width=0.45\columnwidth]{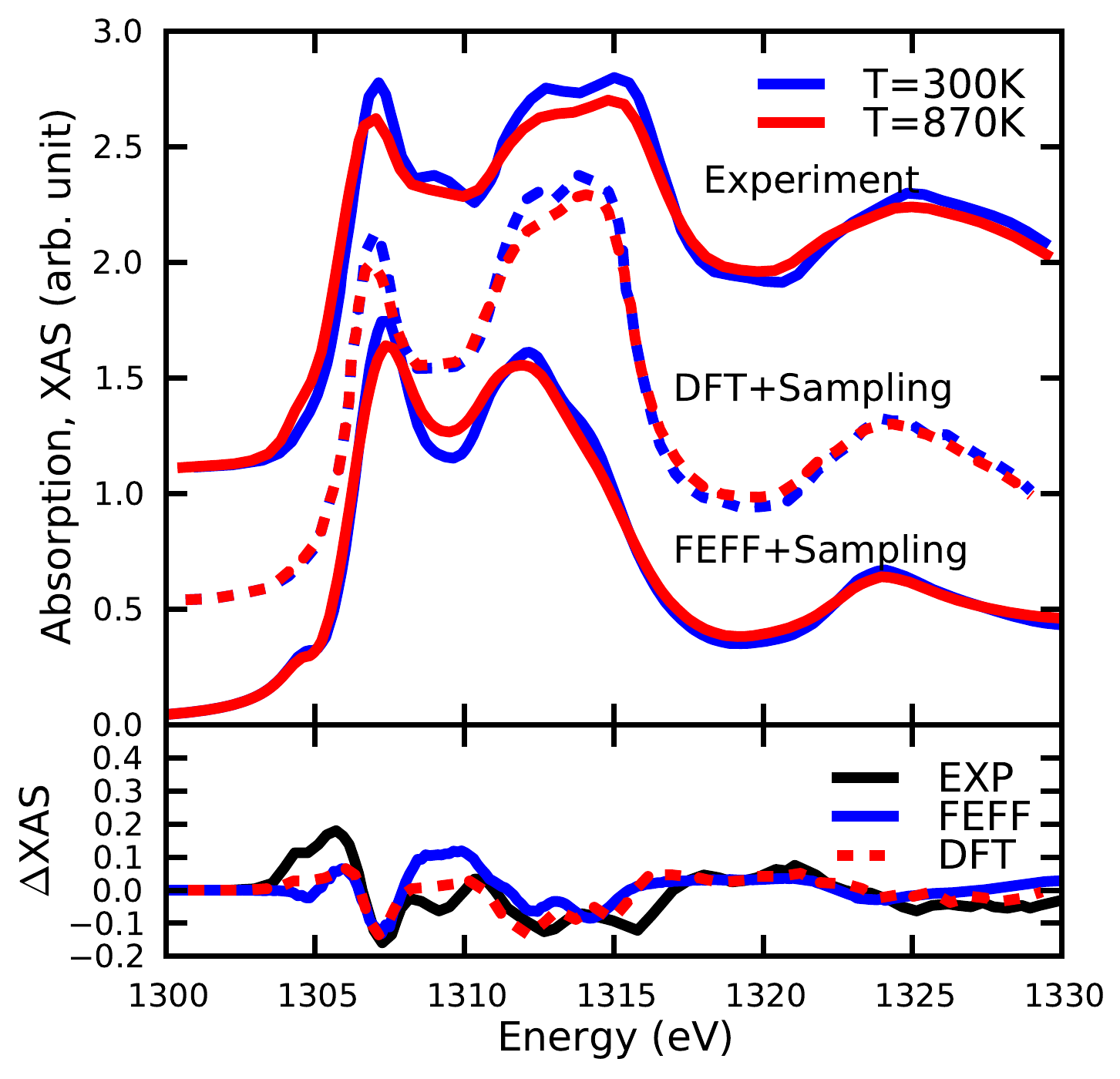}
  \caption{(Top) $\textnormal{K}$-edge XAS of O in equilibrated MgO at T = 300 K and 870 K. The experimental spectra and DFT spectra \cite{Nemausat15} are offset. (Bottom) The spectra difference with respect to T = 300 K.}
    \label{fig:mgo_quasiharmonic}
\end{figure}

The methods described above rely on the quasi-harmonic approximation. At higher temperatures, however, the distribution of path-lengths becomes non-harmonic, and is not well described by standard 
Debye-Waller factors. In addition, at low energies (in the near-edge region)
the use of EXAFS Debye-Waller factors does not capture the effects of
symmetry breaking, even at low temperatures. For these situations
it is preferable to treat disorder using an ensemble average,
with structural snapshots taken from molecular dynamics or vibrational
Monte Carlo sampling.
This method is illustrated in Fig. \ref{fig:mgo_quasiharmonic} for
magnesium oxide (MgO)
with Monte Carlo sampling.
These methods have been implemented within Corvus, and can be
run with a single simplified input file.

\subsection{Multi-electron excitations \label{sect:mbconv}}

Inelastic losses in core level x-ray spectra arise from many-body excitations,
leading to broadening and damping as well as satellite peaks in
x-ray photoemission (XPS) and x-ray absorption (XAS) spectra. 
While calculations of these effects pose a formidable challenge,
there has been significant progress. In particular, the development
of cumulant Green's function methods and the quasi-boson approximation
permit approximate calculations of these effects.
Here we summarize the main results.
Formally the inelastic losses can be partitioned into intrinsic, extrinsic
and interference terms. 
The intrinsic losses arise from excitations due to the sudden creation 
of the core-hole, and can be represented in terms of the core-hole
Green's function $G_c$ (see \cite{Kas16}, Eq. 13).
Calculations of $G_c(t)$ in
the time-domain are facilitated by the cumulant expansion, using the
the Langreth form \cite{Kas16}.
The calculations can be carried out using
real-time time-dependent density functional theory.
The extrinsic losses are obtained from the GW approximation of the photoelectron self-energy and the interference terms are approximated. These effects are
included in the spectra using a convolution with an energy dependent
particle-hole spectral function.

The quasiparticle XAS $\mu_{qp}(\omega)$ is calculated
using FEFF10 as described in Sec.\ \ref{sect:rsgf} with an appropriate
approximation for the FT self-energy.  An approximation for many-body XAS
including the effects of intrinsic excitations is given by a convolution
with the core spectral function $A_c(\omega)$ as in Eq.\ (\ref{eq:mumb}).
More generally, a more complete calculation is given by a convolution
with the particle-hole spectral function, which includes excitations caused by the core-hole (intrinsic excitations),
those caused by the photoelectron (extrinsic excitations), as well as interference between them \cite{KRC}. We have found that several simple approximations are useful: 1) Use of the intrinsic spectral function alone is a fast approximation that is quite good in many cases over the range of energies that includes the near edge, especially in insulators \cite{woicik1,woicik2}.
This is included in FEFF10;
 2) A simple model for the interference and extrinsic can be used for metals, with a single free parameter describing the interference amplitude;\cite{Kas16}; 3) The entire quasi-boson excitation spectrum (intrinsic, extrinsic, and interference) can be modeled from the intrinsic spectrum alone, as the shape of the extrinsic excitation spectrum is, to a good approximation, the same as that of the intrinsic \cite{sky}.

Here we focus on a straightforward extension to the atomic approximation in
FEFF10 for the core-hole spectral function. This is
based on the cumulant expansion approximation for the core-hole
Green's function given by an exponential expression in time,
\begin{equation}
  G_{c}(t) = e^{-i\epsilon_c t + C(t)},
\end{equation}
where $\epsilon_c$ is the core-level quasiparticle energy, and $C(t)$
is the cumulant, which encapsulates all many-body excitations. The
spectral function is given by the Fourier transform of the Green's
function, $A_c(\omega) = -(1/\pi) {\rm Im}\,  G_c(\omega)$.
Within linear response,
the cumulant is related to the density fluctuations caused by the
sudden appearance of the core-hole,
\begin{align}
  C(t) &= \int \frac{d\omega}{\pi} 
  \frac{\beta(\omega)}{\omega^2}\left[e^{-i\omega t}+i\omega t -1\right], \nonumber \\
  \beta(\omega) &= \omega \int dt \Delta(t) e^{i\omega t}, \nonumber \\
  \Delta(t) &= \int d^3r v_c(r)\delta \rho({\bf r},t).
\end{align}
In the above, $\delta \rho(t)$ is the density induced by the sudden
appearance of the core-hole at time $t=0$, $v_c(r)$ is the core-hole
potential, which we approximate as a Coulomb potential centered on the
absorbing atom, and $\beta(\omega)$ can be interpreted as the
quasi-boson excitation spectrum. We calculate the response to the
core-hole within real-time TDDFT using the RT-SIESTA
code \cite{takimoto2007, vila2010, KRC}.

\begin{figure}
\begin{center}
  \includegraphics[width=0.8\columnwidth]{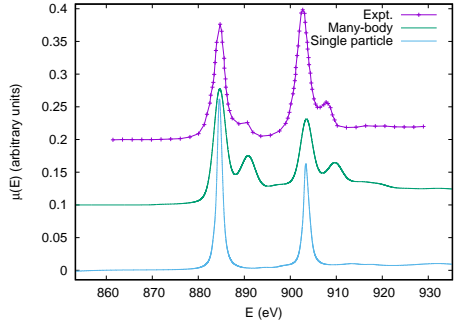}
\caption{Many-body calculation of the Ce M$_{45}$ XANES of CeO$_2$ (green) compared with the experimental EELS data (purple) \cite{ceo2exp} and the single particle calculation (blue).}
\label{fig:mbxanes}%
\end{center}
\end{figure}

This approach has been developed as a workflow in Corvus. The
workflow consists of first calculating the real-time response to the
sudden appearance of the core-hole using a modified version of the
RT-SIESTA code\cite{KRC}. This calculated real-time response is then used
in the cumulant expansion approach to obtain the core-hole spectral
function. Finally, many-body excitation effects are added to the
quasiparticle XAS via convolution with the core-hole spectral
function. Within Corvus, we have implemented a workflow that
allows the user to request the many-body XAS. Starting from a
crystallographic information file (CIF), and a small amount of
additional information provided by the user (Fig. \ref{fig:corvusconv}),
Corvus
then produces input for RT-SIESTA and FEFF using Pymatgen \cite{pymatgen,pymatgenurl}, runs
RT-SIESTA to obtain the core-hole response, and calculates the many-body
core-hole spectral function. FEFF is then used to calculate the
quasiparticle XAS, and finally, Corvus produces the many-body spectrum
by convolving the resulting XAS with the cumulant spectral function. 
Results of this workflow for the $M_{45}-$edge XANES of CeO$_2$ are
shown in Fig.~\ref{fig:mbxanes}, along with the experimental EELS data
\cite{ceo2exp} and the single particle calculation (without many-body convolution). Note the appearance of the satellite peaks at $\sim 890$ eV and $910$ eV in the many-body calculation, in reasonable agreement with those in the experiment. The discrepancy in the M$_4/$M$_5$ ratio between the calculated and experimental results reflect the lack of any treatment of the mixing of the M$_{45}$ holes in the theory. There are various methods for treating this, including multiplet methods, TDDFT, or the solution of the Bethe-Salpeter equation. However, we leave the treatment of these effects to the future.

\begin{figure}
\begin{lstlisting}[language=sh,%
                   frame=single,%
                   basicstyle=\linespread{1.0}\scriptsize\ttfamily,%
                   commentstyle=\color{RoyalBlue}%
                  ]
# General corvus input
# 
# Define the target property
target_list { mbxanes }
# Define which codes to use for this calculation.
usehandlers { helper Feff PyMatGen Siesta phsf mbconv }
# Define structure (cif + supercell)
cif_input{ CeO2.cif }
supercell.dimensions{2 3 4}
# Define absorbing atom type
absorbing_atom_type{ Ce }
# Siesta input
#
# Define how to run in parallel
siesta.MPI.CMD { mpirun }
siesta.MPI.ARGS{ -n 4 }
# SCF convergence parameters
siesta.MaxSCFIterations{ 300 }
siesta.TD.NumberOfTimeSteps{ 200 }
siesta.TD.TimeStep{ 0.5 }
# Define basis (DZP) + d states
# for Ce.
siesta.Block.PAO.Basis{
O           2
 n=2   0   2
   3.305      2.479
   1.000      1.000
 n=2   1   2 P   1
   3.937      2.542
   1.000      1.000
Ce          3
 n=6   0   2 P   1
   8.286      8.183
   1.000      1.000
 n=4   2   2
   0.0        0.0
   1.0        1.0
 n=4   3   2
   3.369      2.148
   1.000      1.000
}
# Define broadening for spectral function.
phsf.broadening{ 1.0 }
# FEFF input
# How to run in parallel
feff.MPI.CMD{ mpirun }
feff.MPI.ARGS{ -n 6 }
# Which edges to calculate
feff.edge{ M4 M5 }
# Use no core-hole
feff.corehole{ none }
# Set energy grid
feff.egrid{
e_grid -10 10 0.05
k_grid last 5 0.05 }
\end{lstlisting}
\caption{Typical Corvus input file for the spectral function convolution.}
\label{fig:corvusconv}
\end{figure}

\subsection{Analysis of x-ray spectra}

We have also implemented a method for analysis of XANES, XES, or similar spectra
using non-linear least squares fitting routines provided by the
lmfit library \cite{newville_lmfit}.
The user requests a fit as the target property, and provides basic
information, including the target of the fit (XANES, XES, or XPS). Free parameters along with their initial values are also defined
in the input (Fig. \ref{fig:corvusfit}). At present, only a few key free parameters are
available for use within Corvus. These include parameters for edge
alignment, Fermi energy adjustment, broadening, overall coordinate
expansion, and rigid movement of clusters of atoms (rigid movement of
a ligand system for example) along a bond.

\begin{figure}
\begin{lstlisting}[language=sh,%
                   frame=single,%
                   basicstyle=\linespread{1.0}\scriptsize\ttfamily,%
                   commentstyle=\color{RoyalBlue}%
                  ]
# General corvus input
#
# Set target property as a fit.
target_list { fit }
# Set the fit target as XES.
fit.target { feffXES }
# Set which handlers to use (FEFF10, lmfit).
usehandlers { fit Feff }
# LMFIT input
#
# Let lmfit know where the experimental data is.
fit.datafile { exp.dat }
# Define fit parameters.
# broadening   - Extra lorenzian broadening beyond core-hole broadening
# delta_e0     - Overall energy shift of spectrum
# amplitude    - Overal scaling factor of spectrum.
# delta_efermi - Shift of Fermi energy cutoff.
# expansion    - Expansion of all coordinates.
# bond         - Shift a cluster of atoms in direction of a particular bond.
fit.parameters {
   broadening    0.0
   delta_e0      -3.7
   amplitude     1.0
   delta_efermi  0.0
   expansion     1.0
   bond 0.0
}
# Define the bond, and cluster of atoms for the "bond" fit parameter. In this
# case, we are only moving the apical nitrogen atom in the direction of the
# absorbing Mn atom. 
fit.bond{ 1 3 }
# FEFF input
# Define how to run in parallel
feff.MPI.CMD{mpirun}
feff.MPI.ARGS{-n 6}
# Define the edge
feff.edge{ K }
# Define the absorbing atom
absorbing_atom{ 1 }
# XES - standard to set no core-hole.
feff.corehole{ none }
# Define the grid.
feff.egrid{
e_grid -45 10 0.25
}
# Define the system using a cluster (xyz like structure).
cluster {
   Mn 0.000000 0.000000 0.000000
   O 0.199422 0.594958 1.791970
   N -0.731983 -1.757497 0.528479
   O -1.808293 0.994578 -0.167048
   N -0.008891 -0.580607 -2.000021
   N 0.953112 1.767136 -0.798793
   N 2.122474 -0.734026 -0.129083
   C 0.979701 -1.713087 -2.091138
   C -0.734189 1.094658 2.558873
.
.
.
}
\end{lstlisting}
\caption{Typical Corvus input file for XANES fitting.}
\label{fig:corvusfit}
\end{figure}

\begin{figure}
  \includegraphics[width=0.45\columnwidth,clip]{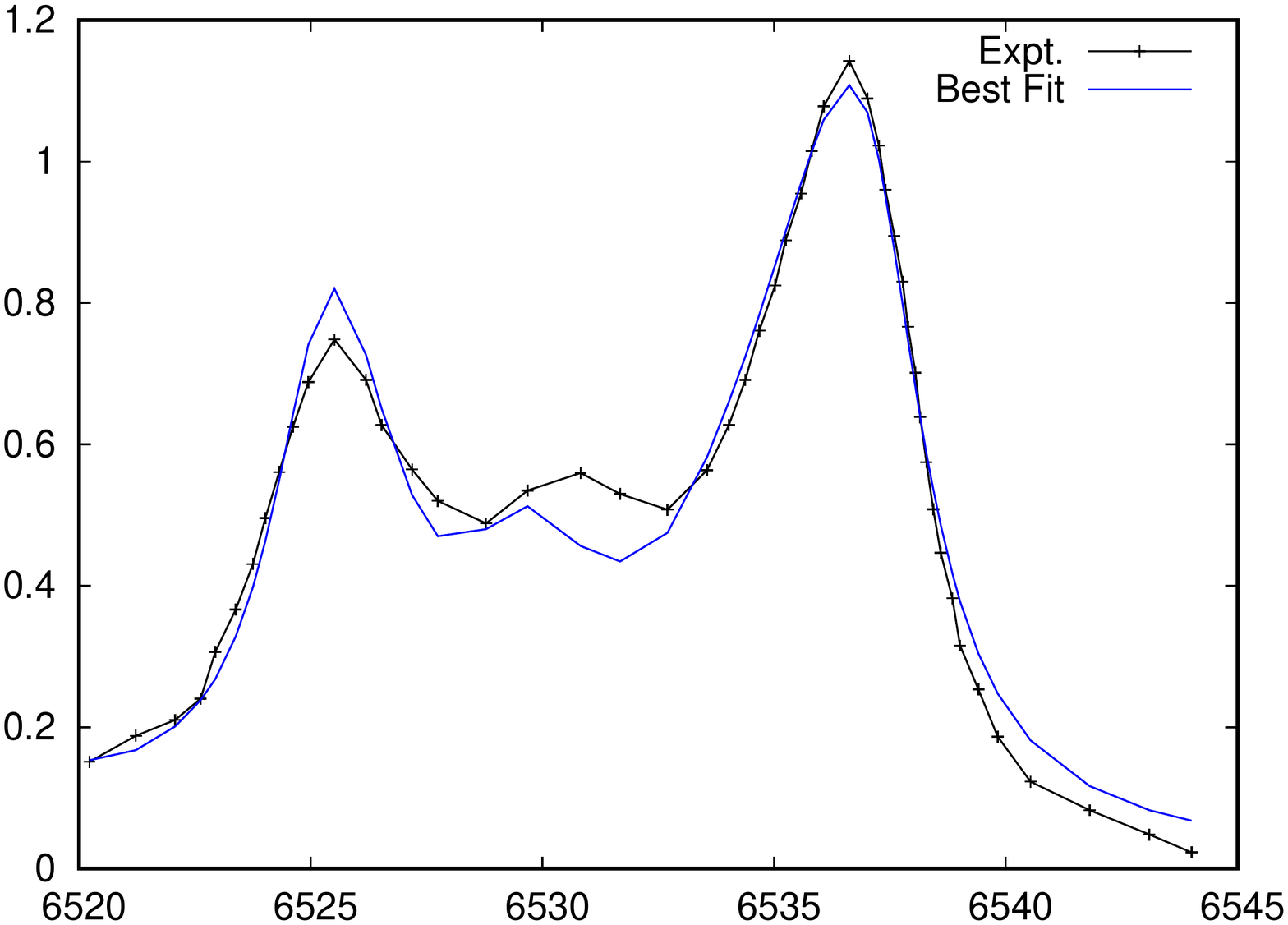}
  \includegraphics[width=0.45\columnwidth,clip]{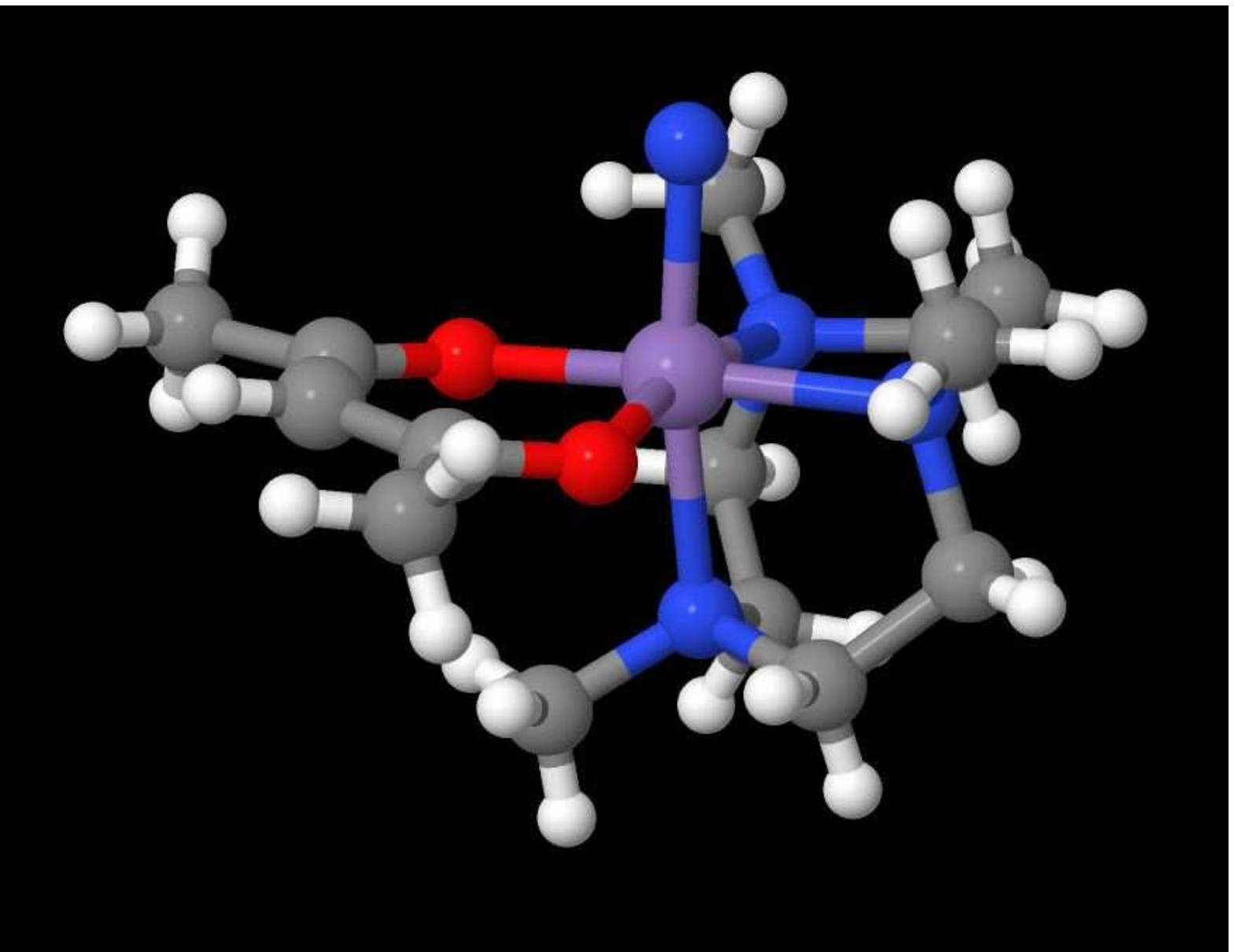}
\caption{Best fit of the FEFF10 calculated Mn K$-\beta$ XES spectrum of ${\rm [LMn(acac)N]BPh}_4$ compared to the experimental data (left). The structure of the molecule (without BPh$_4$) is also shown (right).}
\label{fig:bestfit}
\end{figure}

Figure \ref{fig:bestfit} shows results of a fit to the experimental XES of ${\rm [LMn(acac)N]BPh}_4$ \cite{Smolentsev2009}, along with the ${\rm [LMn(acac)N]}^+$ ion used in the calculations. The main physical parameters used in the fit were the bond-length of the apical nitrogen, and an overall expansion of the molecule. The most important physical parameter is the bond length of the apical nitrogen, which is found to be $1.52 \pm 0.02$ \AA, in good agreement with experimental data ($1.518 \pm 0.004$) \cite{Niemann1996}.

\subsection{Optical constants from UV-VIS to X-Ray}
Optical constants are important for materials design as they describe
the frequency dependent interaction between light and matter
(dielectric constant), or charged particles and matter (energy loss
function). Here we briefly describe the use of FEFF10 and Corvus 
to produce optical constants over a wide range of energies from the
the UV-VIS to X-ray regimes.
The optical constants can all be obtained
from the the imaginary part of the dielectric constant
$\epsilon_2(\omega)$. In particular, the real part of the dielectric
constant can be found via a Kramers-Kronig transform,
\begin{equation}
  \epsilon_1(\omega) = 1 + \frac{2}{\pi}{\mathcal P}\int_{0}^{\infty}\ d\omega'
  \frac{\omega'\epsilon_2(\omega')}{\omega^2-\omega'^2}.
\end{equation}
From the complex dielectric function, various other optical constants
are obtained,
such as the complex index of refraction $n+i\kappa$, absorption
coefficient $\mu$, reflectivity $R$, and energy loss spectrum $L$,
\begin{align}
  &n(\omega)+i\kappa(\omega) = \epsilon(\omega)^{1/2}, \nonumber \\
  &\mu(\omega) = 2\omega/c \kappa(\omega), \nonumber \\
  &R(\omega) = \frac{[n(\omega)-1]^2 +
  \kappa(\omega)^2}{[n(\omega)+1]^2 +
  \kappa(\omega)^2}, \nonumber \\
  &L(\omega) = -{\rm Im}[\epsilon(\omega)^{-1}].
\end{align}
The imaginary part of the dielectric function can be split into
contributions from the valence electrons $\epsilon_{2}^{(v)}(\omega)$, and contributions from the
core electrons $\epsilon_{2}^{(c)}(\omega)$. The core contributions can
be calculated using the standard methods implemented within FEFF, as
described elsewhere and in 
Sec.~\ref{sect:rsgf} \cite{opcons,mpopcons}. Corvus facilitates
these calculations by setting up the correct energy grids, calculating
all edges in the system, including the near-edge and extended regimes,
and summing to obtain the complete contribution from all
core-levels. The valence contribution is more difficult to obtain, but
an approximate method is given by a convolution of the low energy edges
with the appropriate
angular momentum projected densities of states and can give quite
reasonable results. However, in some cases, especially
when strong excitonic behavior is expected, such as in Si, the
Bethe-Salpeter equation must be solved in order to obtain good
results \cite{AI2NBSE}. For such cases we suggest the use of the OCEAN
code, which combines DFT with the NIST Bethe-Salpeter
equation solver \cite{ocean}.

\subsection{Superheavy elements up to {\bf $Z=138$}}

\begin{figure}
\begin{center}
  \includegraphics[width=0.45\columnwidth]{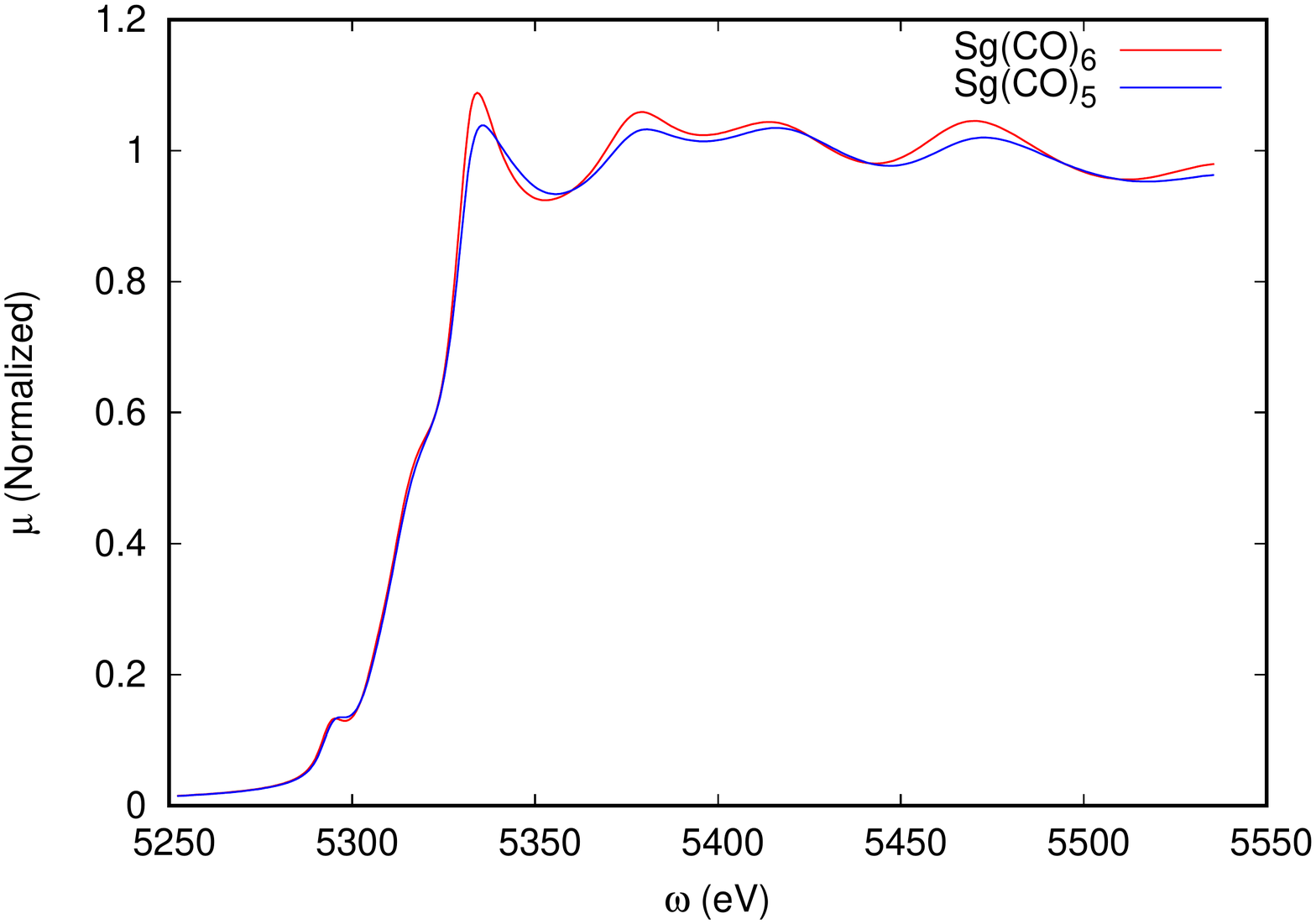}
  \includegraphics[width=0.45\columnwidth]{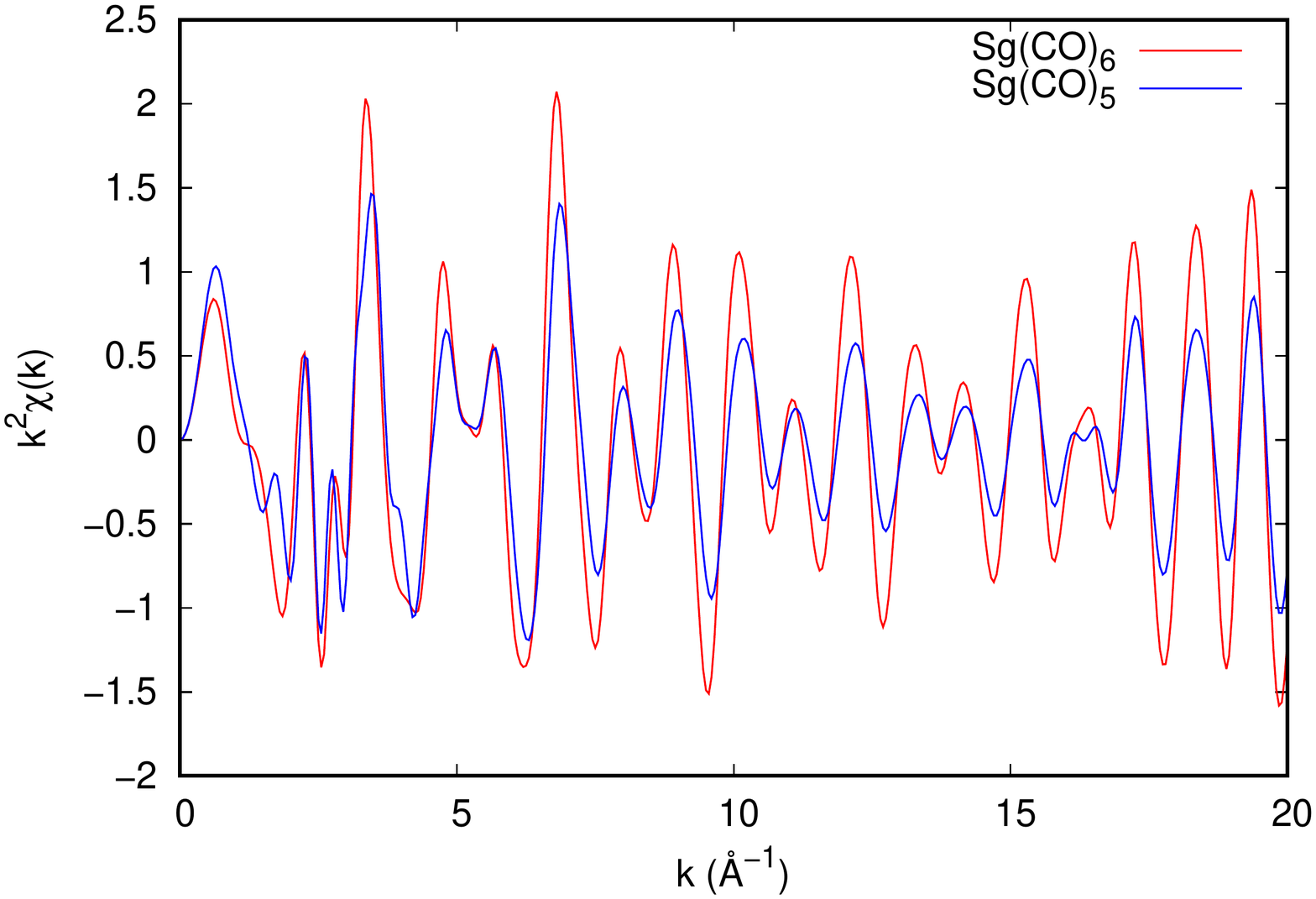}
\caption{M$_5$ XANES (left) and EXAFS (right) of Sg(CO)$_6$ (red)
compared with that of Sg(CO)$_5$ (blue). The change in the
fine-structure reflects the change in the first shell coordination.}
\label{735510}
\end{center}
\end{figure}

Recent experimental and theoretical work has focused on the formation of compounds including superheavy elements \cite{Eichler_2016, Oleynichenko_2018,Ilia__2017,Even_2014,Hammou_2019}. For example, a Seaborgium molecule  ($Z=106$)
Sg$($CO$)_6$ has been detected in gas phase. Since the lifetime of such molecules is only on the order of seconds, experimental methods of investigating their chemical and structural properties must be fast. X-ray spectroscopy is a particularly good candidate, given its ultra-short probe time, element specificity, and ability to probe short-range
order (Fig. \ref{735510}).  As noted above, FEFF10 now includes the capability to treat
systems with superheavy elements up to $Z$= 138 \cite{Zhou2017},
based on an automated single configuration version of the multi-configurational
Dirac-Fock atomic code \cite{DESCLAUX1973311,Ankudinov_1996}.
These calculations can be carried out, for example, with a FEFF-only
Corvus workflow, or with more elaborate calculations as desired.

\subsection{Other developments}

Together with the workflows described above, we have developed several other  
workflows that simplify calculations based on FEFF10. These include
for example, sums over edges (e.g., L$_{23}$ or M$_{45}$) and
configurational averaging, i.e., averaging of spectra over the unique sites in a
crystal, and calculations of resonant inelastic x-ray scattering (RIXS). 

\section{Summary and Conclusions \label{sect:conclusion}}

We have developed FEFF10, a new version of the RSGF code FEFF, together with an efficient workflow tool to enable advanced calculations of x-ray spectra by combining FEFF10 with auxiliary codes. This hybrid approach permits
  a number of extensions and features 
including finite-temperature with auxiliary calculations of structure
and vibrational properties.  The approach is generally applicable
to systems throughout the periodic table, temperatures
up to the WDM regime, and non-equilibrium systems including simulations
of time-resolved pump-probe experiments.
The code   utilizes an improved Dirac-Fock code with the capability of treating super-heavy elements up to $Z=138$.
We have also described the augmentation of FEFF10 with auxiliary codes
through automated workflows implemented within the Python-based
Corvus workflow manager.  This coupling facilitates automated advanced
calculations. The calculations have been compared to both 
experiment and other theoretical methods, and generally give good results.
Many extensions and applications are possible. 

\begin{acknowledgements}

The development of FEFF10 and Corvus software was carried out within the
Theory Institute for Materials and Energy Spectroscopies (TIMES) at SLAC,
and is supported by the U.S. DOE, Office of Basic Energy Sciences, Division of Materials Sciences and Engineering, under contract
DE-AC02-76SF00515. The development of the finite-temperature extension
in FEFF10 was supported by the DOE Office of Science BES Grant
DEFG02-97ER45623, with computational support from NERSC, a
DOE Office of Science User Facility, under Contract No. DE-AC02-05CH11231.

\end{acknowledgements}

%



\bibliography{FEFF10}
\end{document}